\documentclass{hnp06}

\newcommand{\beq}{\begin{eqnarray}}
\newcommand{\eeq}{\end{eqnarray}}

\newcommand{\del}{\partial}

\def\Slash#1{/\hspace{-0.23cm}{#1}} 
\def\ms{M_S} \def\mq{m_q} \def\ma{M_A}

\begin{document}
\setcounter{page}{1}

\title{Roper resonance in a quark-diquark model}

\author{Keitaro Nagata and 
Atsushi Hosaka}

\address{Research Center for Nuclear Physics (RCNP)\\
Osaka University, Ibaraki 567--0047, Japan}

\maketitle

\abstracts{
We discuss a new description for the Roper resonance, 
the first nucleon excited state of $J^P = 1/2^+$, 
in a model of strong diquark correlations.
Treating the scalar-isoscalar and axial-vector--isovector diquarks 
as independent degrees of freedom, 
two states having nucleon quantum numbers
are constructed.  
Due to the scalar and axial-vector nature of the diquarks, the 
two nucleon states have different  internal structure of spin and isospin.  
This yields the mass splitting of order several hundreds MeV, 
and hence the two states are identified with the nucleon and Roper.  
We demonstrate this scenario in a simple two channel problem.
}

%================================================
\section{Introduction}
%================================================

The first excited states of baryons having the same
spin and parity as the ground states, $J^P = 1/2^+$, are
known experimentally in various $uds$ flavor sectors.   
In particular the nucleon resonance $N(1440)$ which is 
called the Roper resonance has been investigated extensively.  
In the naive quark model of harmonic oscillator potential, 
it is assigned as the nodal excitation of $2\hbar \omega$, 
whose excitation energy is 
twice as large as that of the $l= 1$ negative parity state of 
$1\hbar \omega$~\cite{Isgur:1978xj,Isgur:1979be}.  
In experiments, however, these two states of $1/2^{\pm}$
are almost degenerate, or more precisely, 
the Roper appears slightly lower than the $1/2^-$ state.  
This feature is not only in the nucleon channel but 
also in almost all light flavor channels~\cite{Takayama:1999kc}.  

Because of its low mass, the Roper resonance has been 
considered as a collective excitation of the nucleon.  
Majority of such description resorted to the monopole 
excitation of the ground state~\cite{Brown:1983ib,Hatsuda:1986ph,Breit:1984tf,Hayashi:1984bc,Zahed:1984qv,Mattis:1984ak,Hosaka:1988pj}.
Another interesting collective picture was  proposed 
by one of us and others, where the baryon resonances 
were described as collective rotational states of a deformed 
intrinsic state~\cite{Takayama:1999kc}. 
This scheme may explain masses of not only the Roper but also almost 
all baryon resonances with a few parameters.  

Yet another interesting idea was proposed by 
Weinberg~\cite{Weinberg:1969hw} and later 
considered by Beane and 
collaborators~\cite{Beane:2002ud,Beane:2002td}, 
where the Roper was regarded as a chiral partner 
together with the nucleon and delta resonance.  
Such a view is interesting since their 
properties are directly related to chiral symmetry of QCD 
with its spontaneous breaking.  
The model we consider here has some relevance to this 
approach, although its precise relation is not yet fully 
explored.  
However, the use of explicit diquark structure of two quarks 
is convenient when discussing chiral symmetry transformation 
properties of baryons~\cite{nagata_prep}.  

With the above considerations, 
we study the nucleon and Roper resonance in a 
quark-diquark model with scalar and axial-vector diquarks 
treated explicitly.  
Our model set up is simple in which the nucleons
are regarded as bound states of a quark and a diquark 
through a contact interaction with suitable regularization.  
The bound state problem is then treated in a path integral 
formalism.
The gap equations for the nucleon states 
are then obtained, which are 
equivalent to the non-relativistic Schr\"odinger equations.  
We consider a coupled channel problem for the two states of 
the scalar diquark and axial-vector diquark nucleons.  
Then their linear combinations are 
regarded as the physical nucleon and Roper after the diagonalization 
of the Hamiltonian.  
We discuss the masses and possible spatial structure of the 
nucleon and Roper resonance.  

%================================================
\section{Model}
%================================================

%---------------------------
\subsection{Diquarks}
%---------------------------

The basic assumption is the diquark correlation in the 
nucleon.  
The relevance of diquarks in recent hadron spectroscopy 
has been discussed by Jaffe~\cite{Jaffe:2005md,Jaffe:2004ph}.  
Due to its maximal attractive interaction as shown in Table~\ref{diquark}, 
the scalar diquark is expected to play major role for nucleon structure.   
In practice,
another axial-vector diquark is also important.   
If the two diquarks are regarded as independent
degrees of freedom in a three-quark baryon 
system, the two nucleon states can emerge as independent 
states having the ground state spatial configuration.  
Such a possibility is not allowed in the SU(6) quark model, 
where one of the two states is forbidden due to the Pauli principle.

\begin{table}[htdp]
\caption{Matrix elements of color-spin $V_{CS}$ and spin-flavor $V_{SF}$ 
interaction in various diquark channels.  The interactions are
$V_{CS, SF} \sim 
- \sum_{ij}\frac{\lambda^a(i)}{2} \frac{\lambda^a(j)}{2} 
\; \frac{\vec \sigma(i)}{2} \cdot \frac{\vec \sigma(j)}{2}$, 
where the $\lambda$ matrices are for either color or flavor space.}
\begin{center}
\begin{tabular}{c c c c c}
\hline
$qq(C,S,F)$ & $(\bar 3_C, 0_S, \bar 3_F)$ & $(\bar 3_C, 1_S, \bar 6_F)$
    & $(\bar 6_C, 0_S, \bar 6_F)$ & $(\bar 6_C, 1_S, \bar 3_F)$ \\
    & $D_S$ & $D_A$ & & \\
\hline
$V_{CS}$ & $-1/2$ & $1/6$& $1/4$ & $-1/12$ \\
$V_{SF}$ & $-1/2$ & $-1/12$ & $1/4$ & $1/6$ \\
\hline
\end{tabular}
\end{center}
\label{diquark}
\end{table}%

In the quark model language, the scalar and axial-vector
diquarks have spin-isospin structures as 
\beq
D_S &\sim& 
\frac{1}{\sqrt{2}}(\uparrow \downarrow - \downarrow \uparrow)
\; 
\frac{1}{\sqrt{2}}(ud -du) \, , \nonumber \\
D_A &\sim& 
\frac{1}{\sqrt{2}}(\uparrow \downarrow + \downarrow \uparrow)
\; 
\frac{1}{\sqrt{2}}(ud +du) \, , 
\label{diquarks}
\eeq
where arrows express spin up and down states and $u,d$ the flavor 
$u, d$ quarks.  
The two diquarks are combined with another quark 
to make two basis states for the nucleon and Roper:
\beq
N_S = D_S q\, \; \; N_A = D_A q \, , 
\eeq 
where in $N_A$, the proper combination should be made for 
spin and isospin to take $J = I = 1/2$.  
In the SU(6) quark model, only the sum of equal weight is allowed
for the nucleon:
\beq
N _{\rm QM}= \frac{1}{\sqrt{2}}(N_S + N_A) \, .
\eeq

If the two diquarks are active degrees of freedom, then we will have 
in addition to the two nucleon states of $N_S$ and $N_A$, the delta
$\Delta(1232)$ as described by the combination of the 
axial-diquark and a quark.  
Therefore, we would be lead to the idea that 
the three states ($N_A$, $N_A$ and $\Delta$) may be 
described on the same footing as a family of quark-diquark states.  
This reminds us of the chiral model for the 
these particles~\cite{Weinberg:1969hw,Beane:2002ud,Beane:2002td}.
At present, the relation of the two descriptions is not 
clear, but it would be interesting to investigate the properties 
of the quark-diquark baryons with chiral symmetry.   

%---------------------------
\subsection{Lagrangian}
%---------------------------

Having the quark and diquark fields, we write down an 
SU(2)$_L\times$ SU(2)$_R$ chiral quark-diquark 
model~\cite{Abu-Raddad:2002pw,Nagata:2003gg,Nagata:2004ky}, 
\begin{eqnarray}
{\cal L} &=& \bar{\chi}_c(i\rlap/\del - m_q) \chi_c +
D^\dag_c (\del^2 + M_S^2)D_c
\nonumber \\
&+&
{\vec{D}^{\dag\;\mu}}_c 
\left[  (\del^2 + M_A^2)g_{\mu \nu} - \del_\mu \del_\nu\right]
\vec{D}^{\nu}_c +{\cal L}_{int},
\label{lsemibos}
\end{eqnarray}
where $\chi_c$, $D_c$ and $\vec{D}_{\mu c}$ are 
the constituent quark, scalar diquark and axial-vector diquark fields
with color index $c$, 
and $m_q$, $M_S$ and $M_A$ are their masses. 
The axial-vector diquark carries the Dirac and isospin indices, 
since it is a spin one isovector particle.  
In Eq.~(\ref{lsemibos}), the quark $\chi$ is the constituent 
quark of non-linear representation.  
Therefore, chiral symmetry is preserved 
in the presence of the constituent quark mass $m_q$.

%The diquarks are expressed conveniently 
%by the quark-bilinears\cite{Abu-Raddad:2002pw}, 
%$D_c\sim \epsilon_{abc}\tilde{\chi}^b\chi^c,\ \vec{D}_{\mu c}\sim \epsilon_{abc}\tilde{\chi}^b \gamma_\mu \gamma_5 \vec{\tau}\chi^c$.
%Here $\tilde{\chi}=\chi^T C\gamma_5 i\tau_2$ and $\epsilon_{abc}$ is
%the antisymmetric tensor.

\begin{figure}[ht]
\centering
\includegraphics[width=9cm]{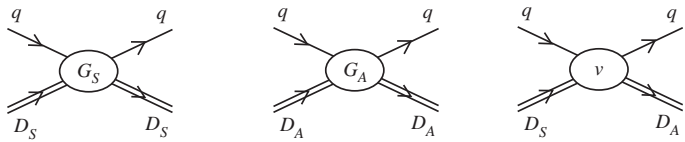}
\caption{Quark-diquark interactions in various channels.}\label{interaction}
\end{figure}

The interaction term ${\cal L}_{int}$ includes the diagonal and non-diagonal 
(mixing) parts as given by
\beq
{\cal L}_{int}
&=&
G_S\bar{\chi}_cD^\dagger_c D_{c^\prime}\chi_{c^\prime}
+ G_A\bar{\chi}_c\gamma^\mu\gamma^5
\vec{\tau}\cdot\vec{D}^\dagger_{\mu c}
\vec{\tau}\cdot\vec{D}_{\nu c^\prime}\gamma^\nu\gamma^5 \chi_{c^\prime}
\nonumber \\
&+& v(\bar{\chi}_cD^\dagger_c\gamma^\mu\gamma^5
\vec{\tau}\cdot\vec{D}_{\mu c^\prime} \chi_{c^\prime}+\bar{\chi}_c\gamma^\mu\gamma^5
\vec{\tau}\cdot\vec{D}^\dagger_{\mu c}
D_{c^\prime}\chi_{c^\prime})\, , 
\label{eq:twoc}
\eeq
where $G_S$ and $G_A$ are the coupling constants for the quark and scalar
diquark, and for the quark and  axial-vector diquark, respectively.  
The coupling
constant $v$ causes the mixing between the scalar and axial-vector
channels (see Figure~\ref{interaction}).

%figure---------------------------------------------
%figure---------------------------------------------

%---------------------------
\subsection{Hadronization}
%---------------------------

Introducing the
auxiliary fields for baryons, we can rewrite the Lagrangian as 
(omit the color indices for brevity)
\begin{eqnarray}
{\cal L} &=& \bar{\chi}(i\rlap/\del - m_q) \chi \;+\;
D^\dag (\del^2 + M_S^2)D
\nonumber \\
&+&
{\vec{D}^{\dag\;\mu}} 
\left[  (\del^2 + M_A^2)g_{\mu \nu} - \del_\mu \del_\nu\right]
\vec{D}^{\nu}+\bar{\psi}\hat{G}\psi-\bar{B}\hat{G}^{-1}B\, , 
\label{L_inter}
\end{eqnarray}
where
$B=(B_1, B_2)^T$ is a two component auxiliary baryon field, 
whose components 
correspond to  scalar and axial-vector channels; $B_1\sim D\chi$ 
and $B_2\sim \vec{\tau}\cdot\vec{D}_\mu\gamma^\mu\gamma^5\chi$.
In Eq.~(\ref{L_inter}) we have introduced matrix notations of $2 \times 2$
as
\begin{eqnarray}
\psi&=&\left(\begin{array}{c} D\chi\\ 
\vec{D}_\mu\cdot\vec{\tau}\gamma^\mu\gamma^5\chi\end{array}\right),\quad
\bar{\psi}=
\left(\begin{array}{cc} 
\bar{\chi}D^\dagger, & 
\bar{\chi}\vec{D}_\mu^\dagger\cdot\vec{\tau} \gamma^\mu \gamma^5 \end{array} \right),\\
\hat{G}&=&\left(\begin{array}{cc} G_S & v \\ v & G_A \end{array}\right).
\end{eqnarray}

In the hadronization
procedure\cite{Abu-Raddad:2002pw,Cahill:1988zi,Reinhardt:1989rw}, the
quark and diquark fields are eliminated and an effective meson-baryon Lagrangian
is obtained in the $\mbox{tr}\log$ form as
\begin{equation}
{\cal L}=-\bar{B}\hat{G}^{-1}B+i\mbox{Tr}\ln(1-A).
\label{eq:trln}
\end{equation}
Here the matrix $A$ is defined by
\begin{eqnarray}
A&=&\left(\begin{array}{cc} a_{11} & a_{12}\\ a_{21}& a_{22}\end{array}\right),\\
a_{11}&=&\Delta^T \bar{B}_1 S B_1,\\
a_{12}&=&\Delta^T\bar{B}_2 \tau^i \gamma^\nu\gamma^5  S B_1,\\
a_{21}&=&(\Delta_{\rho\nu}^{lj})^T \bar{B}_1 S \gamma^\mu\gamma^5\tau^j  B_2,\\
a_{22}&=&(\Delta_{\rho\nu}^{lj})^T \bar{B}_2\gamma^\nu\gamma^5\tau^i S \gamma^\mu\gamma^5\tau^j B_2.
\end{eqnarray}
where $S$, $\Delta$ and $\Delta_{\mu\nu}$ are the propagators of the
quark, scalar diquark and axial-vector diquark, respectively.

Expanding the tr log formula in powers of meson and baryon fields,  
we obtain the self-energies of the nucleons as
\begin{eqnarray}
{\cal L}=
\bar{B}\left(\begin{array}{cc} 
\Sigma_S(p) & 0 \\ 0 & \Sigma_A(p)\end{array}\right)B 
- \frac{1}{|\hat{G}|}\bar{B}\left(\begin{array}{cc} G_A 
& -v \\ -v & G_S \end{array}\right) B,
\label{eq:lag2}
\end{eqnarray}
where $|\hat{G}|=\mbox{det} \hat{G}=G_SG_A-v^2$.
The loop integrals of the self-energies are given by 
\begin{eqnarray}
\Sigma_S(p)&=&-i N_c\int\frac{d^4k}{(2\pi)^4}
\frac{1}{k^2-\ms^2}\frac{\Slash{p}-\Slash{k}+\mq}{(p-k)^2-\mq^2},\\
\Sigma_A(p)&=&-iN_c\int\frac{d^4k}{(2\pi)^4}
\frac{k^\mu k^\nu/\ma^2-g^{\mu\nu}}{k^2-\ma^2}
\nonumber \\
& & \; \; \; \times \; 
\delta_{ij}\gamma_\nu\gamma_5\tau_j
\frac{\Slash{p}-\Slash{k}+\mq}{(p-k)^2-\mq^2}\tau_i\gamma_\mu\gamma_5 . 
\label{eqn:selfa}
\end{eqnarray}
Here $N_c$ is the number of colors. 
In our computation, 
the divergent integrals of  (\ref{eqn:selfa}) are regularized
by the three momentum cutoff scheme\cite{Nagata:2004ky}.

%---------------------------
\subsection{Diagonalization}
%---------------------------

We consider nucleon properties in the center of mass 
system of the nucleon.  
The self-energies $\Sigma_S$ and $\Sigma_A$ are then 
expanded in powers of the four momentum in the rest frame
$p_\mu=(p_0,\ \vec{0})$, 
\begin{eqnarray}
\Sigma_S(p_0)-\frac{1}{|\hat{G}|}G_A&=&Z_S^{-1}(p_0\gamma^0-a_S),
\label{SigmaS}\\
\Sigma_A(p_0)-\frac{1}{|\hat{G}|}G_S&=&Z_A^{-1}(p_0\gamma^0-a_A)
\, .
\label{SigmaA}
\end{eqnarray}
\label{Sigma}
The bare baryon fields $B_{1,2}$ are then renormalized as
\begin{eqnarray}
\left(\begin{array}{c} B_1 \\ B_2 \end{array}\right)
=\left(\begin{array}{c}\sqrt{Z_S} B_1^\prime\\
\sqrt{Z_A} B_2^\prime\end{array}\right) ,
\end{eqnarray}
with which the Lagrangian (\ref{eq:lag2}) can be written as
\begin{equation}
{\cal L}=\bar{B}^\prime(p_0\gamma^0 - \hat M )B^\prime\, ,
\end{equation}
where the mass matrix $\hat{M}$ is given by
\begin{eqnarray}
\hat{M}=\left(\begin{array}{cc} a_S & -\sqrt{Z_S Z_A}
\frac{v}{|\hat{G}|}\\ -\sqrt{Z_S Z_A} \frac{v}{|\hat{G}|} & a_A
\end{array}\right).
\label{eq:mmatrix}
\end{eqnarray}

Now the mass matrix can be diagonalized through 
a unitary transformation:
\begin{eqnarray}
B^\prime&=&U^\dagger N\, , \; \; \; 
U\hat{M}U^\dagger
=\left(\begin{array}{cc} M_1 & 0 \\ 0 & M_2 \end{array}\right).
\end{eqnarray}
One finds
\begin{equation}
{\cal L}=\bar{N_1}(p_0\gamma^0-M_1)N_1+\bar{N_2}(p_0\gamma^0-M_2)N_2,
\end{equation}
where the physical eigenvalues $M_{1,2}$ and eigenvectors 
$N=(N_1,\ N_2)^T$ are obtained as
\begin{eqnarray}
M_{1,2}&=&\frac12\left[a_S+a_A\pm
\sqrt{(a_S-a_A)^2+4Z_S Z_A
\left(\frac{v}{|\hat{G}|}\right)^2}\right]\, , 
\label{det_mass}\\
N_1&=&\cos\phi B_1^\prime+\sin\phi B_2^\prime \, , \; \; 
N_2=-\sin\phi B_1^\prime+\cos\phi B_2^\prime \, , 
\label{eigenfn}
\end{eqnarray}
and the mixing angle $\phi$ is given by 
\begin{equation}
\tan 2\phi=\frac{2\sqrt{Z_S Z_A}v}{(a_A-a_S)|\hat{G}|}
\end{equation}
Eq. (\ref{det_mass}) should be read as a self-consistent 
equation where the quantities on the right hand side 
are functions of $M_{1,2}$.  
The equations are then equivalent to the Schrodinger 
equation for the quark-diquark system interacting through 
the delta function type interaction with suitable cutoff.  

%================================================
\section{Results and discussions}
%================================================

First let us fix model parameters. 
The constituent masses of $ud$ quarks $m_q$ and the three momentum 
cutoff $\Lambda$ are fixed in such a way that they 
reproduce meson properties in the NJL model\cite{Vogl:1991qt,Hatsuda:1994pi}. 
The masses of the diquarks may be calculated in the NJL
model~\cite{Vogl:1991qt,Cahill:1987qr}, 
but here we treat them as parameters.  
In this way we employ $m_q$=390 MeV, $\Lambda$=600 MeV, 
$M_S$=650 MeV and $M_A$=1050 MeV.  
The mass difference $M_A - M_S$
may be related to that of the nucleon and delta.  
In the quark-diquark model, the delta is described as a bound state 
of an axial-vector diquark and a quark, while the nucleon is 
expected to be dominated by the scalar diquark component.  
Hence, we expect that the $N\Delta$ mass difference is 
roughly given by the mass difference 
of the axial and scalar diquarks.  
This qualitatively justifies the mass difference 
$M_A - M_S \sim 400$ MeV that we adopt.  

\begin{figure}[ht]
\centering
\includegraphics[width=5.5cm]{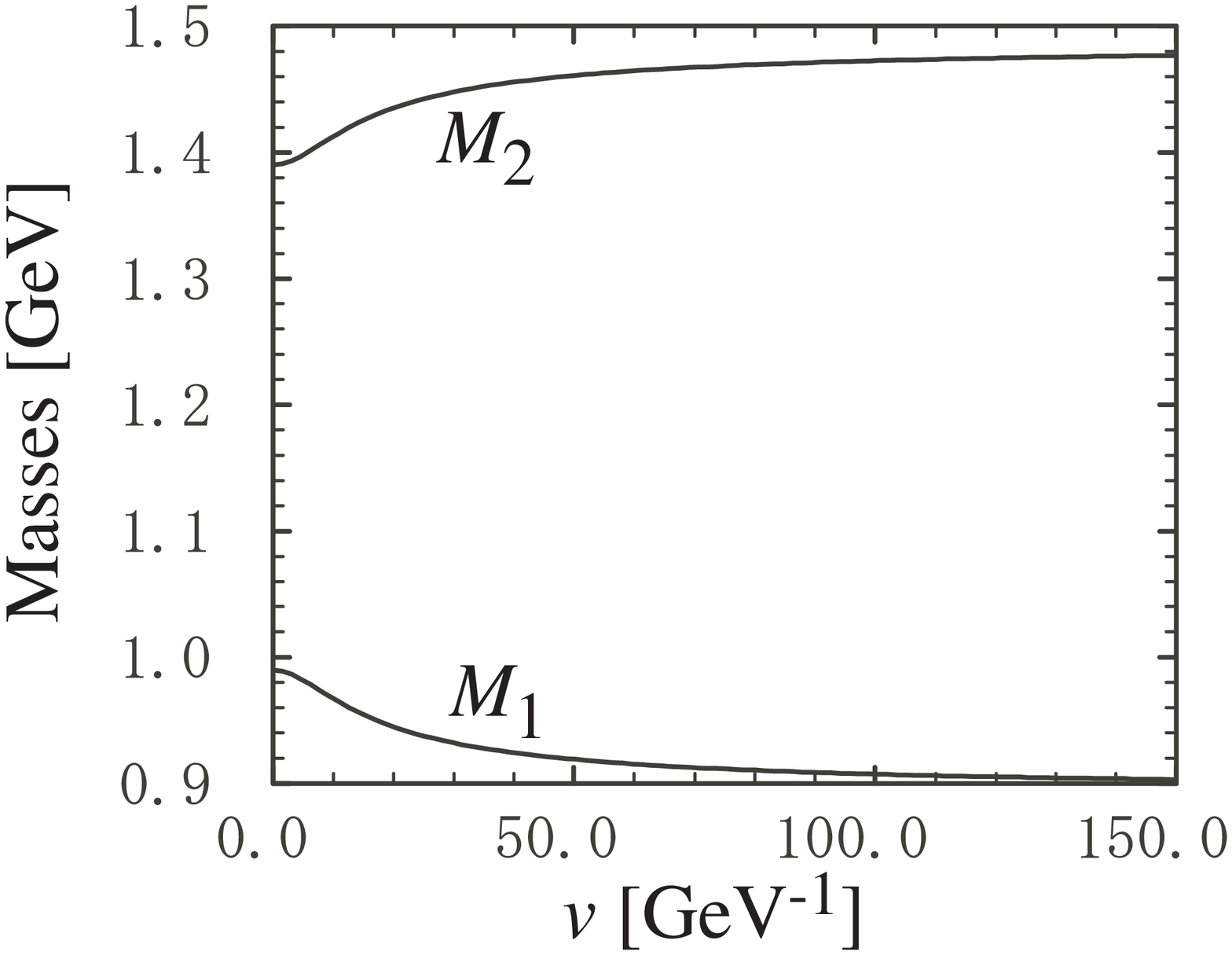}
\includegraphics[width=5.5cm]{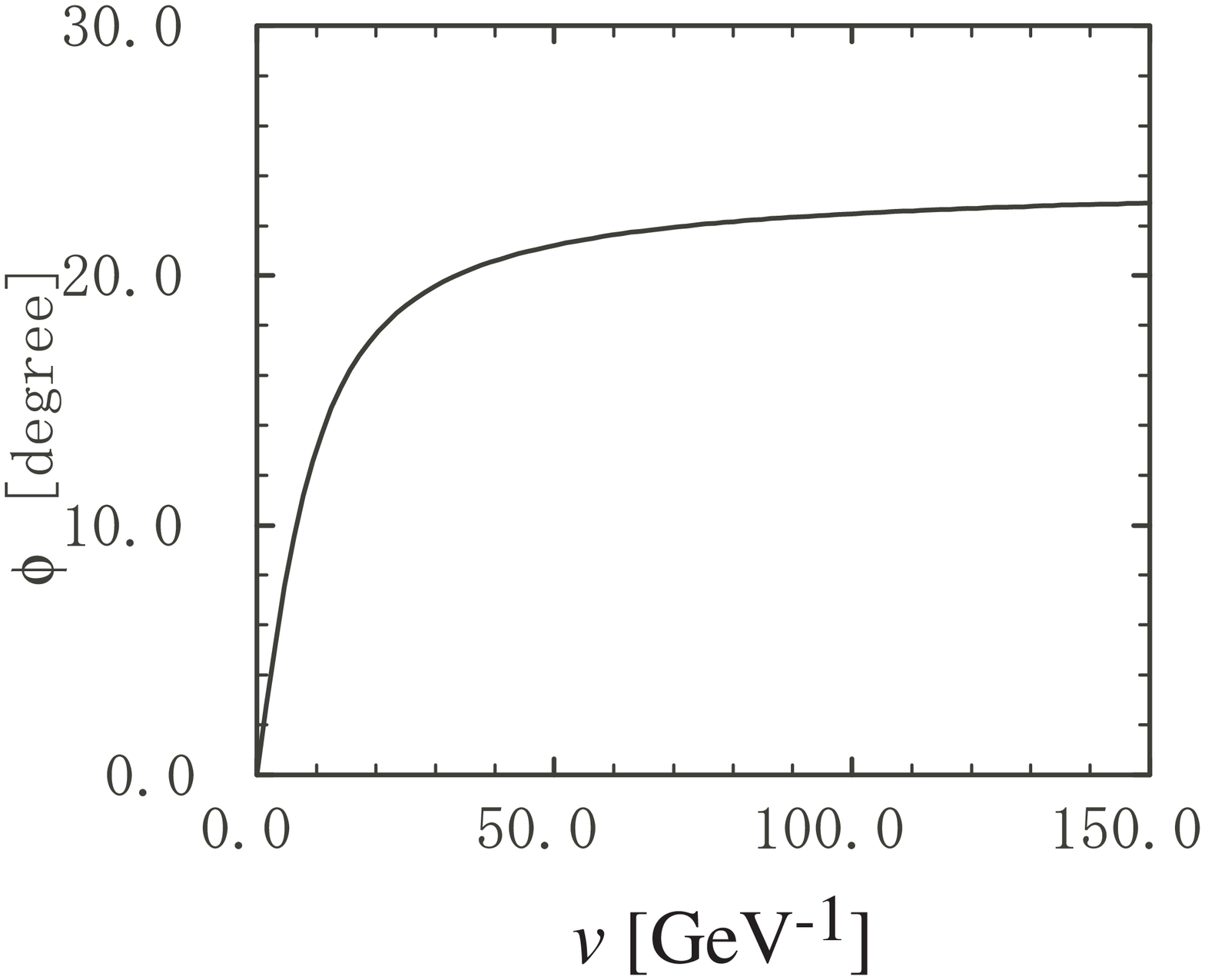}
\caption{The masses $M_{1,2}$ as functions of the 
mixing strength $v$ (left) and the mixing angle $\phi$ (right).  
The coupling constants $G_S$ and $G_A$ are determined such 
that the binding energies of the quark-diquark system takes 
50 MeV.}\label{fig:mass2}
\end{figure}

The masses of the two nucleon states may be studied 
as functions of the coupling constants, $G_S$, $G_A$
and $v$.  
For instance, we can fix the diagonal strengths $G_S$, $G_A$, 
and vary the off-diagonal strength $v$ to see the effect of 
the mixing.  
In this procedure, however, the binding energy of the 
quark-diquark system also changes which significantly 
affects the sizes of the nucleon and Roper.  
For some coupling constants, the nucleon becomes unphysically 
too large.  
In order to overcome this difficulty, we fix the quark-diquark 
binding energies to be always 50 MeV for both scalar and 
axial-vector diquark nucleons as $v$ is varied.  
This can be realized by choosing $G_S$ and $G_A$
appropriately, and the resulting bound states 
produces nucleon size 
reasonably~\cite{Abu-Raddad:2002pw,Nagata:2003gg}.   
The results for the masses of the nucleon and Roper 
are shown in Figure~\ref{fig:mass2}.  
We find that $M_1=$ 0.94 GeV, $M_2$=1.44 GeV 
and $\phi$ =18 degree at
$v\sim$ 22 GeV$^{-1}$.
Due to the fixed binding energy, we find that the plot 
looks very much the same as the one familiar in a 
two level problem of the quantum mechanics.  

%figure---------------------------------------------
%figure---------------------------------------------

The present identification of the Roper resonance is 
very much different from the conventional picture; 
in the quark model, it is described as an excited 
state of $2\hbar \omega$ with 
$(n,l) = (1,0)$, where $(n,l)$ are the principle and 
angular momentum quantum numbers of the harmonic oscillator
wave function.  
The excitation energy of such a state is as high as 
$2 \hbar \omega \sim 1$ GeV for the oscillator parameter 
$\omega \sim 0.5$ GeV, and many mechanisms have been 
proposed to lower the energy~\cite{Glozman:1995fu}.  
In the present picture the two nucleons are 
described as quark-diquark bound states, but 
with different diquarks of scalar and axial-vector ones.  

In the quark model, these diquarks correspond to 
the $\rho$ and $\lambda$ type two-quark states, 
which in the limit of SU(6) spin-flavor symmetry
can not be independent degrees of freedom due to the Pauli
principle when constructing the nucleon state of $J^P = 1/2^+$.  
In the present case, if the 
strong correlation between the quarks 
is at work, the two quark states violate the 
SU(6) symmetry, and they can be independent.  
In the harmonic oscillator basis, 
the two quarks of the diquarks are in the ground state 
but with being correlated.  
The energy difference is therefore supplied not by the 
difference in the single particle energies of nodal excitation, 
but rather by the residual correlation between the quarks.  
This is the mechanism that makes the mass of the 
Roper significantly lower than the conventional 
radial exitation.  

\begin{figure}[ht]
\centering
\includegraphics[width=9cm]{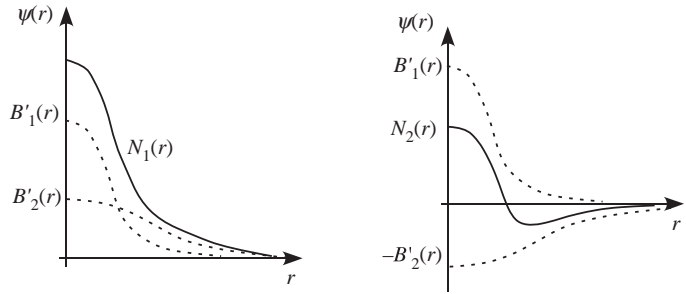}
\caption{A qualitative sketch for quark distributions 
for the nucleon and Roper.}\label{wf_dist}
\end{figure}

It is also interesting to look at the spatial structure of the 
nucleon and Roper when they are given as superpositions of 
the two diquark components (\ref{eigenfn}).  
Intrinsically, the scalar diquark is a tightly bound state of two 
quarks while the axial-vector looser. 
Therefore, the wave function of 
the scalar diquark nucleon $B_1^\prime$ is more 
compact than the axial-vector diquark nucleon
$B_2^\prime$.  
If the wave functions of $B_1^\prime$ and $B_2^\prime$
are coherently added with a constructive phase for the 
nucleon ($N_1$), then for the Roper ($N_2$) they are 
added destructively.  
Hence, we expect qualitatively different structure in the 
wave function for the nucleon and Roper 
as shown in Figure~\ref{wf_dist}.  
It is interesting that the wave function of the Roper 
has a nodal structure as a consequence of the 
two components of $B_1^\prime$ and $B_2^\prime$
just as in the nodal excitation in the naive quark model.  

%figure---------------------------------------------
%figure---------------------------------------------

%================================================
\section{Summary}
%================================================

In this report, we have discussed  the nucleon and Roper resonance in
the chiral quark-diquark model. 
It was shown that the two states appear as the ground state in 
orbital configuration but with different spin and isospin structure.
In the SU(6) quark model, one of these states is forbidden due to 
the Pauli exclusion principle, which, however, can survive as
an independent degree due to diquark correlations.  

In a sample calculation, we have reproduced 
the masses of the nucleon and Roper by employing 
suitable parameters.  
This encourages us to further investigate 
this picture for the nucleon and Roper.  
As a straightforward application, the present model should be 
tested for electromagnetic properties of the nucleon and 
the Roper.  
For the nucleon, we expect that the inclusion of the 
axial-vector diquark improves the small magnetic moments 
and the axial-vector coupling constants when only 
scalar diquark is considered.  

The description of the nucleon and Roper as the ground state 
with different spin-isospin structure reminds us the 
Weinberg's idea that they are regarded as chiral partners 
which belong to the same chiral multiplet.  
In this way we may be able to explain the masses and 
the coupling relations among the nucleons and the Roper.  
Relation with the chiral symmetry is particularly interesting 
and will be studied when written the diquark fields 
explicitly by the quark fields.  
Some consequences of such descriptions 
will be reported elsewhere~\cite{nagata_prep}.

%===============================================

\end{document}